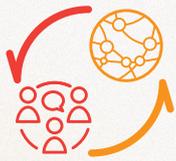

# INDCOR

INDCOR WHITEPAPER 4

# Evaluation of Interactive Narrative Design for Complexity Representations

**EDITORS**
Christian Roth, Breanne Pitt, Lāsma Šķestere, Jonathan Barbara

**AUTHORS**
Ágnes Karolina Bakk, Kirsty Dunlop, Maria del Mar Grandio, Miguel Barreda, Despoina Sampatakou, Christian Roth, Breanne Pitt, Lāsma Šķestere, Jonathan Barbara, Michael Schlauch

**CONTRIBUTORS**
Nicole Basaraba, Noam Knoller, Dennis Haak

COST
EUROPEAN COOPERATION
IN SCIENCE & TECHNOLOGY

# INDCOR

**EDITORS**
Christian Roth, Breanne Pitt, Lāsma Šķestere, Jonathan Barbara

**AUTHORS**
Ágnes Karolina Bakk, Kirsty Dunlop, Maria del Mar Grandio,
Miguel Barreda, Despoina Sampatakou, Christian Roth,
Breanne Pitt, Lāsma Šķestere, Jonathan Barbara, Michael Schlauch

**CONTRIBUTORS**
Nicole Basaraba, Noam Knoller, Dennis Haak

**INDCOR WHITEPAPER 4**

# Evaluation of Interactive Narrative Design for Complexity Representations

This article/publication is based upon work from COST Action INDCOR, CA18230, supported by COST (European Cooperation in Science and Technology).

COST (European Cooperation in Science and Technology) is a funding agency for research and innovation networks. Our Actions help connect research initiatives across Europe and enable scientists to grow their ideas by sharing them with their peers. This boosts their research, career and innovation.

www.cost.eu

Funded by the European Union

# Table of Contents

# Abstract

While a strength of Interactive Digital Narratives (IDN) is its support for multi-perspectivity, this also poses a substantial challenge to its evaluation. Moreover, evaluation has to assess the system's ability to represent a complex reality as well as the user's understanding of that complex reality as a result of the experience of interacting with the system. This is needed to measure an IDN's efficiency and effectiveness in representing the chosen complex phenomenon. We here present some empirical methods employed by INDCOR members in their research including UX toolkits and scales. Particularly, we consider the impact of IDN on transformative learning and its evaluation through self-reporting and other alternatives.

**VISION**

WG3 aims to gain a holistic and transdisciplinary understanding of the moment-to-moment, short term, and long-term outcomes of engagement with interactive narratives for complexity representation.

**MISSION**

WG3 will evaluate the impact of IDN complexity representations in audiences, in a manner that productively informs the other work groups and different stakeholders. Specifically, we identify and establish key success indicators, collect and develop methods and frameworks to assess and evaluate interactive narratives, as well as analyze the impact of different aspects and manifestations of interactive narratives representing complex topics.

**DEFINITION**

IDNs are narrative experiences that can be changed by an audience and which are created for the digital medium. What can be changed (for example outcome, progression, perspective) varies as does the particular form. Popular IDN forms include narrative-focused video games, interactive documentaries, hypertext fictions and VR/AR/MR experiences.

# 01. Introduction

Evaluation of Interactive Digital Narrative (IDN) aims at answering the question of what makes a successful IDN experience. IDNs offer interactors the opportunity to explore a simulated narrative space, often inhabited by believable characters that they can interact with. IDNs hold particular promise as a modern learning and entertainment experience, with good stories that engage learners and thus convey content and meaning with interaction enabling active learning through experimentation. Learn more about IDNs in general in INDCOR WP 0 (2023).

One concrete application, for instance, is to broaden the learners' horizon by supporting them in overcoming confirmation bias resulting from algorithmic recommendations that create a filter bubble which gives an alternate picture of reality (see Figure 1).

IDNs can give learners' access to multiple perspectives on a particular issue, helping them understand the relationships between the multitude of factors involved.

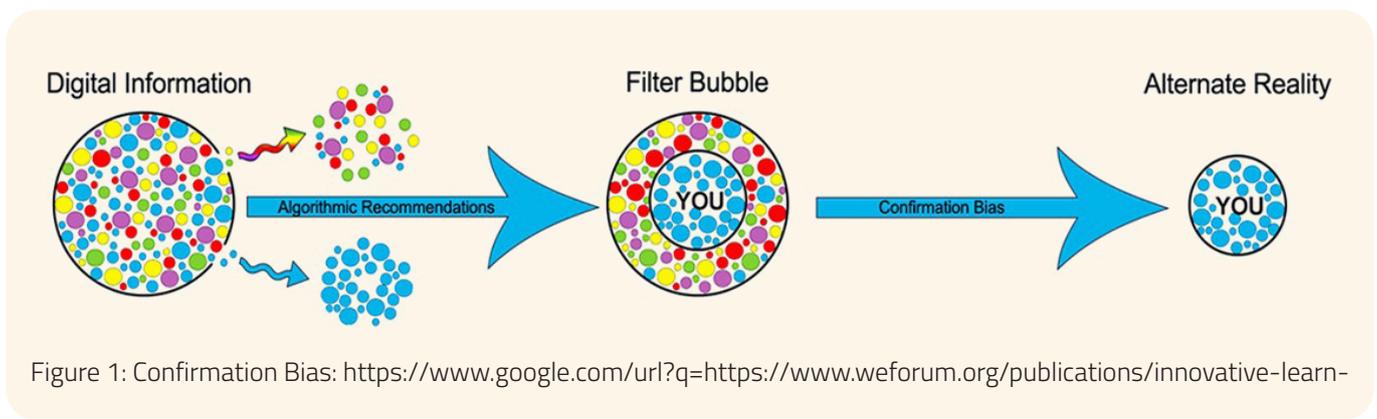

Figure 1: Confirmation Bias: https://www.google.com/url?q=https://www.weforum.org/publications/innovative-learn-

Interactors broaden their knowledge when they know a wider variety of (counter-)arguments (in terms of social actors and epistemological viewpoints), important questions arising, and associated value systems that go beyond their personal views (see Figure 2).

Interactors deepen their knowledge when they are able to go beyond enumerating arguments at a shallow level, in arguing on arguments, and in negotiating the meaning of important underlying concepts (e.g., in the debate on peacebuilding, the concepts of democracy, equity and fairness are important).

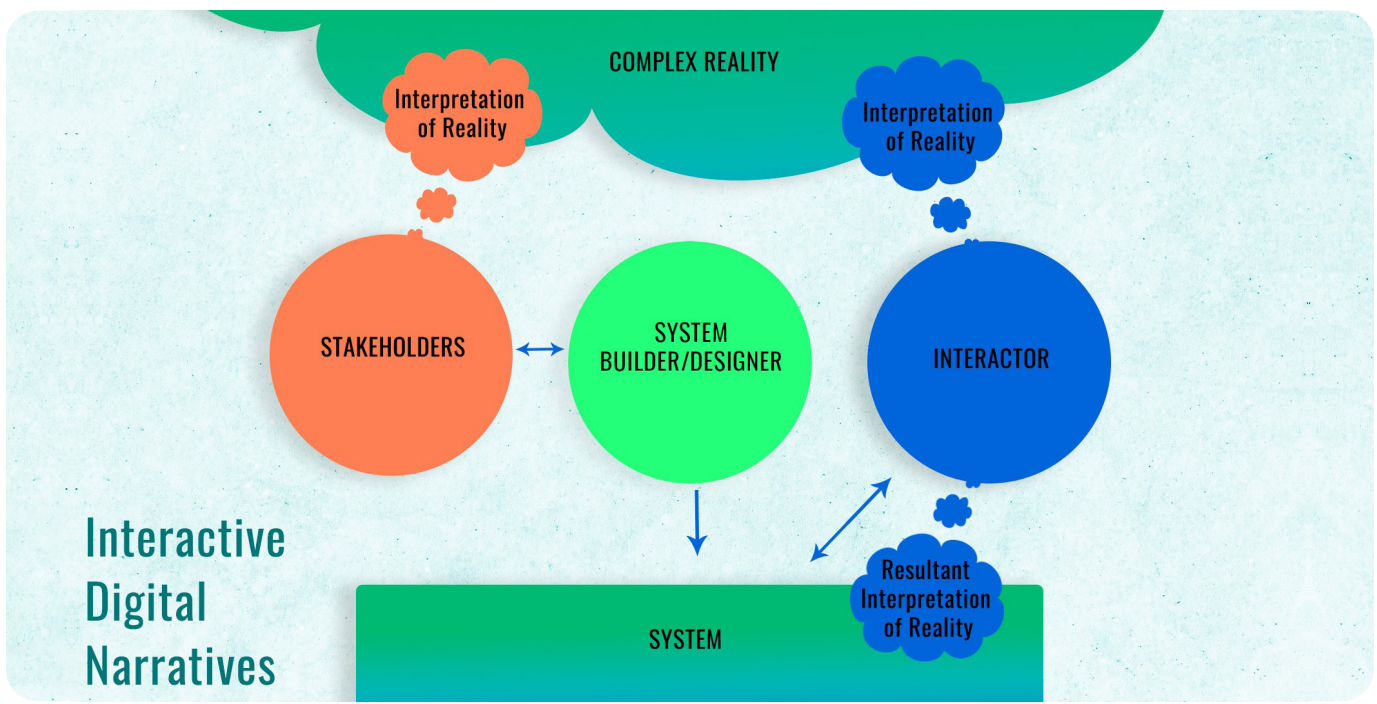

Figure 2: Informing the interactor's interpretation of reality as a result of interacting with the IDN System

## Multiperspectivity

Complex IDN Simulations scaffold learning as learners explore the complexities of conflicting perspectives and pragmatic demands through direct and personal experience (bottom-up or frog's eye view). By repeatedly engaging, failing and trying again, learners develop a systemic overview and understanding of the complex relationships and dynamics between characters (top-down or bird's-eye view).

> **Frog's-eye view (bottom-up)**: Point of view from the ground or from within a system through direct, personal experience and observation, close to life situations and their context. More narrowly focused.

> **Bird's-eye view (top-down)**: Looking down on the system from a high position to see all of it. Allows to see relationships within a hierarchical system consisting of macro units and inter-connected subunits.

Multiperspectivity also brings in ethical demands on the design of IDN. The selection of perspectives being represented in the IDN need to be ethical in their choice and equal in their treatment. Simplification and trivialisation of complex phenomena is unethical towards the represented. While scaffolding interaction to assist interactors in building an understanding of how the outcomes are influenced by their inputs into the system is an ethical approach towards IDN design (Barbara, Koenitz, Bakk, 2021). Find out more about IDN terminology in INDCOR WP 1 (2023).

## Purpose of Evaluation

Evaluation of complexity representations is needed to answer these questions:

- How effective is an IDN at representing complex systems? (Learn more about IDN as complexity representation in INDCOR WP 3 (2023))
- How efficient are IDNs at minimizing authorial bias and encouraging systemic thinking? (Learn more about IDN authoring in INDCOR WP 2 (2023))
- How well is the topic depicted in the IDN / simulation? Does it portray the complexity well?
- How to make the learning with IDNS more engaging and impactful? (Learn more about the impact of IDNs on societal issues in INDCOR WP 5 (2023))
- Which design elements make IDNS work well or not so well?

## Evaluation Techniques

Evaluation involves two interwoven dimensions: the system's ability to represent a complex reality and the user's understanding of that complex reality as a result of the experience of interacting with the system. We distinguish evaluation approaches based on explicit, subjective self-report data (interviews and questionnaires) and implicit, objective measures (physiological, statistics from artifacts/telemetry). Mixed-methods, combining qualitative and quantitative data promise a holistic insight into the user experience.

## Empirical Methods

Empirical methods provide means to gather knowledge based on actual artifacts, as well as players' reactions to them. Both qualitative and quantitative methods have a place in IDN research and should be combined for maximum effect for a holistic understanding of the effects of design methods on the user experience, using explicit, subjective (interviews, questionnaires) and implicit, objective data (physiological measurements, statistics from artefacts). Of equal importance is a clear understanding of what aspect of design is to be evaluated: "As narrative systems grow in their capacities, the community needs a set of well-designed evaluation methods and criteria that can bring insights on the systems as well as the stories they provide" (Zhu, 2012). A majority of studies mainly focused on the global appreciation of the Interactive Narrative Experience (INE), inquiring users about their experience after a play session. For instance Roth's measurement toolbox (2015) aims to provide a shared platform with a uniform perspective for cross-evaluation and benchmarking of INEs produced by different systems (Roth & Koenitz, 2016). Schoenau-Fog (2011) sees the desire to continue the experience as a fundamental requirement of any IDN and developed the Engagement Sampling Questionnaire to investigate engagement as Continuation Desire. Both approaches are combined in the work of Estupiñán Vesga (2020) who aims for a fine-grained analysis by finding unobtrusive moments for user feedback during the experience.

# 02. Evaluation of IDN

In this paper, we discuss two different forms of evaluation: evaluation of IDN systems and evaluation of IDN user experience. The evaluation of an IDN system is understood to be a comparison of the IDN's ability to express the complex system being represented. Following Figure 1 above, an evaluation of an IDN system is carried out on the whole expanse of the complex reality modeled into the system against the stakeholder's interpretation of the complex reality. On the other hand, an evaluation of an IDN user experience is the evaluation of the interactor's usually non-exhaustive exploration of the IDN system and may be in comparison with their original interpretation of that complex reality or with the complex reality modeled into the system, or with the stakeholder's interpretation of the complex reality.

## IDN Systems Evaluation

The assessment of IDN systems is non-trivial because they are not tied to a specific medium. IDNs can be delivered through a combination of modalities (e.g., interactive text, sound, image, animation or video) and these can be expressed in different media such as video games, hypertext fiction, or interactive documentaries for example. Each medium brings with it delivery and interaction mechanisms specific to its form for which a standardized evaluation assessment cannot be applied. The suitability of the chosen medium for the representation of the intended experience is another criterion that challenges evaluation while being a critical factor for the assessment of the IDN artifact's ability to portray the intended complex phenomenon.

Thus evaluating an IDN is a composite process composed of media-specific evaluation techniques, as well as a suitability assessment, for each modality employed to deliver a particular aspect of the IDN.

## IDN User Experience Evaluation

Besides the system designer's ability to portray the stakeholder's interpretation of the complex reality into the system, an IDN, seen as a communication medium, needs to be evaluated from the audience's perspective, particularly in their active role as interactors.

Furthermore, there are a multitude of measurable dimensions of user experience that need to be measured in order to capture the full effect of the IDN on its audience. These require a variety of measuring tools which have been developed by members of this working group and are described hereunder, followed by a focus on IDN's affordance for transformative learning.

## IDN UX Toolkits

Revi, Millard, and Middleton (2020) offer a systematic analysis of user experience dimensions for interactive digital narratives. From our group, Roth's (2015) measurement toolbox is featured in the summary and we will briefly introduce it here as it has been used in a series of IDN user experience evaluation studies. The measurement toolbox is modular and offers short and long scales to measure 11 relevant user experience dimensions.

A key element towards fulfilling the potential of IDNs with complexity representations is the creation of a satisfying user (learning) experience. Roth's approach enables the systematic and quantitative study of IDN user experiences, connecting research in psychology based on Entertainment Theory with a humanities-based perspective. Specifically, Roth and Koenitz (2016) map Murray's influential theoretical framework to Roth's empirical dimensions and thus connect an analytical framework to empirical research.

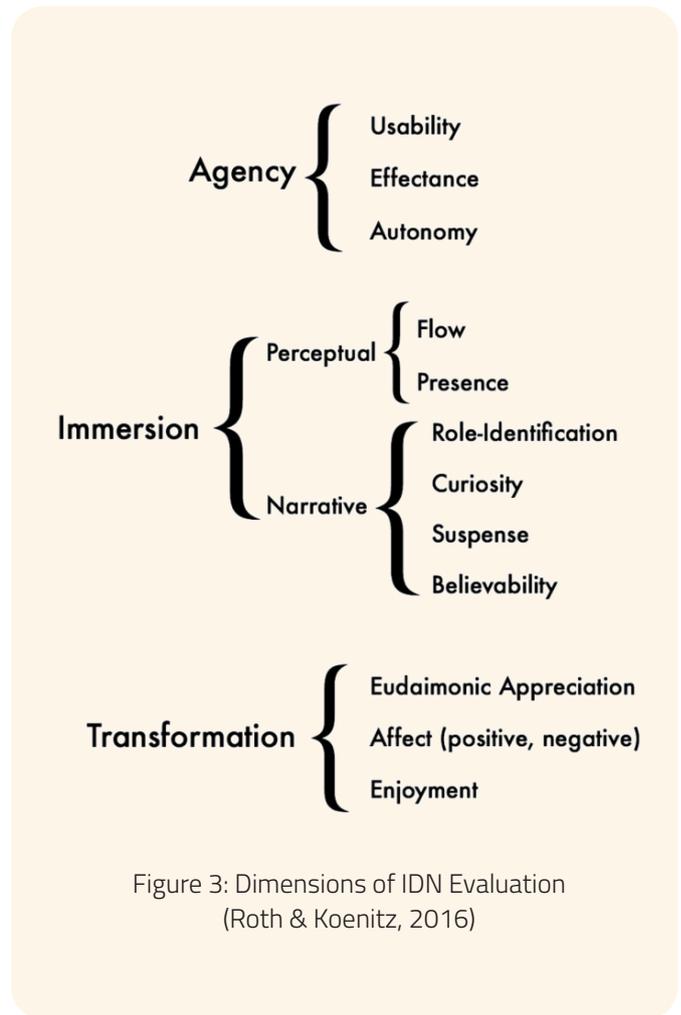

Figure 3: Dimensions of IDN Evaluation (Roth & Koenitz, 2016)

## Narrative Consistency

A complex system has many variables that influence its outcome. Every small variation can cause a great difference in the results. Understanding how these changes generate such huge gaps between the final output goes a long way to understanding the complex system.

Interactive digital narratives allow for the representations of such complex systems, providing narrative outcomes based on the interactor's choices and agency along the narrative. The underlying system cannot be understood through a single interactive session. A second session, with different actions taken along the way, usually does not suffice to understand how all the different variables interact to produce the narrative outcome either. Indeed, any expectations built from the first session will more often than not be broken by the outcome of the second session, and the third, and so forth. Multiple interactive sessions need to be experienced to explore a sufficient number of interactive variations and build an understanding of the relationships between choices made and outcomes achieved. This helps them build a mental model of the complex phenomenon being represented as they map the outcomes to their inputs into the IDN system.

The Narrative Consistency scale (Barbara, 2018) is a 5-point Likert scale (1 conflicting, 2 inconsistent, 3 irrelevant, 4 cohesive, 5 consistent) that can be used to compare narrative experiences based on whether they meet expectations raised in prior engagements. Studies show that the learning effects of pro-social IDNs depend on interactors perceiving their actions as meaningful while having a clear understanding of the consequences. Narrative Consistency can assess the participants' understanding of the modeled social context by measuring how the interactive narrative's outcomes meet their expectations across replays. Should interactors rate the IDN low on the Narrative Consistency scale, i.e. inconsistent or conflicting, then this would reflect their lack of understanding of the underlying complex system. High ratings on the Narrative Consistency scale, such as consistent or cohesive, would indicate a strong understanding of the complex system.

## Case Studies

The following two case studies demonstrate the evaluation of IDN works using the aforementioned measurement toolbox (Roth, 2016). We summarize key takeaways.

### CASE STUDY #1

The Angstfabriek (Dutch for fear factory) is an educational installation in the form of a complete building and lets visitors experience fear-mongering and the related safety industry, with the goal of eliciting reflection, insight and discussion. Interactors undergo a 'transformation as masquerade' by becoming whistleblowers as they see behind the curtain of fear-mongering and the industry benefiting from it.

Roth (2019) evaluated the potentially transformative user experience of this installation via a focus group interview and a pilot user experience study. This revealed the importance of sufficient scripting of visitors regarding their role and agency, highlighting the conceptual connection between IDN design and interactive theater design. Interactors did not know what to expect, what role to inhabit and how to perform it within the unknown rules and constraints of the interactive installation, and thus only a few visitors made use of the disruptive role-playing and barely sabotaged the production of fear-inducing media messages. Reflection on the experience was left up to the visitors to discuss it afterwards.

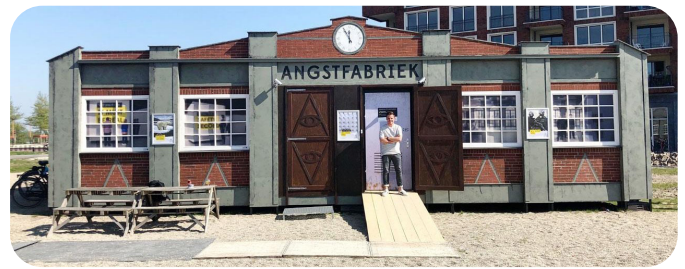

**KEY TAKEAWAYS from Kangstfabrie**

When designing for impact, an expensive one-shot intervention is not enough: include reflection and discussion on the experience as part of the overall design. Allow participants to register for follow up information to increase the potential learning effect. Interactors need to know their role and possible actions via sufficient scripting. This highlights the conceptual connection between IND and interactive theater design.

## CASE STUDY #2

Traditional unilinear narrative accounts of the most popular view - usually of the victor - prevent understanding of the reality of contemporary piracy having so many different perspectives: the hijacker, the hijacked, the negotiators.

The Last Hijack Interactive (2014) is an online interactive documentary which allows readers to choose between different perspectives on piracy in Somalia, that of the ship's captain, his wife, the pirate, and their respective lawyers.

These different perspectives should allow the interactor to build their own representation of the given information and thus better understand such a complex situation.

However, research by Van Enschot et al. (2019) shows that a lack of IDN literacy skills stops the reader from taking full advantage of the medium while underdeveloped perspectives, such as that of the pirate, do not help to deliver the full picture of a complex situation.

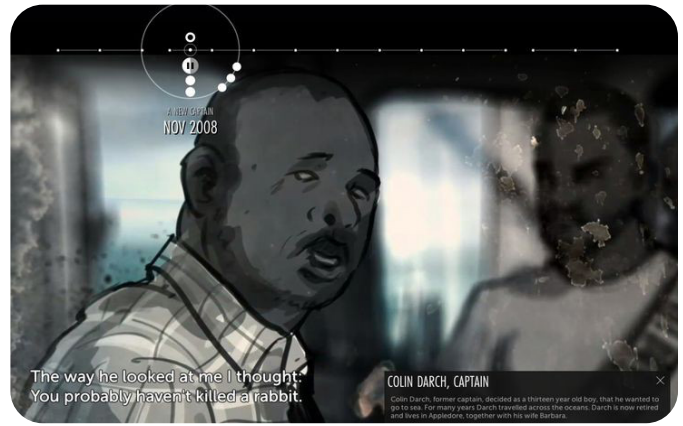

**KEY TAKEAWAYS from Last Hijack Interactive**

Leaving certain perspectives underdeveloped (e.g. the pirate's point of view) can lead to a bias in viewers. Makers have a high responsibility for a fair representation of complex matters. Lacking literacy in optimally using interactive narrative systems hampers the user experience and thus learning effects of a complexity IDN.

# 03. Evaluation of Transformative Learning

**DEFINITION**

Transformation is a fundamental aesthetic quality of interactive narratives with the potential to positively change interactors' thinking and behavior. Understanding complex topics involves different types of transformations and connects to the Transformative Learning Theory.

In her influential book, Hamlet on the Holodeck, Janet Murray (1997) proposes transformation as one of the characteristic pleasures of digital environments and describes three distinct meanings of transformation:

**Transformation as enactment**. The interactive narrative allows the interactor to transform themselves into someone else for the duration of the experience.

**Transformation as variety**. The experience offers a multitude of variations on a theme. The interactor is able to exhaustively explore these kaleidoscopic variations and thus gain an understanding of the theme and different aspects.

**Personal transformation**. The experience takes the player on a journey of personal transformation, potentially altering their attitudes, beliefs, and behavior.

## Personal transformation

Personal transformation denotes the product of the game experience, potentially altering the interactor's attitudes, beliefs, and behavior. Roth and Koenitz (2016) further conceptualized personal transformation via the notion of eudaimonic appreciation, the link that connects the aesthetic presentation of an IDN to a personal dimension rooted in previous experience. Interactors construct personal meaning from engaging with a narrative. This conceptualisation is based on entertainment theory, which differentiates between pleasure-seeking hedonic and truth-seeking eudaimonic motivations to use entertainment media (cf. Oliver & Bartsch, 2010). The exploration of aesthetic content is intrinsically motivated and understanding the meaning of an aesthetic presentation is a challenge that is driven by curiosity and identification, based on personal taste and background. Hence, aesthetic and eudaimonic appreciation is often linked to the personal meaning individuals attach to media offerings (Rowold, 2008).

Empirical studies on emotionally moving and challenging experiences in games (Bopp et al., 2016; 2018) seem to largely support this notion of eudaimonic appreciation. These instances are not solely characterized by their intensely emotional impact, but players also describe these experiences as thought-provoking, personally meaningful, staying with them long after gameplay, as well as often shaping how they see and act in the world. For instance, being confronted with racism against one's character in Skyrim (Carver, 2011), coping with personal loss by mourning the death of one of the titular brothers in Brothers: A Tale of Two Sons (Fares, 2013), or dealing with ethical dilemmas in the Mass Effect series (Watamaniuk, Karpyshyn & Hudson, 2007).

Murray's understanding of personal transformation can be connected to Mezirow's theory, which gives an important insight on how such transformations can form. The **Transformative Learning Theory** developed by Jack Mezirow sees transformative learning as the critical awareness of unconscious suppositions or expectations and the evaluation of their relevance for making an interpretation.

Mezirow suggests that we interpret meaning out of experiences through a lens of deeply held assumptions that experience disorienting dilemmas as their understanding of the world is challenged and contradicted. This reassessment of their assumptions and processes of meaning-making results in a transformative experience that shifts the learner's unconditional acceptance of available information into a conscious and reflective way of learning that supports real change. This perspective transformation results in changes in one's understanding of the self and lifestyle as well as a revision of their belief system (Mezirow, 1991).

This type of learning transcends simply acquiring knowledge as it concerns deep, useful and constructive learning which offers constructive and critical ways for learners to consciously give meaning to their lives.

The **Reflection Framework** by Fleck and Fitzpatrick (2010) drew from Mezirow and is intended for Human-Computer Interaction. There is considerable overlap between their individual reflection stages and Murray's notions of transformation: I.e., descriptive reflection denotes reasoning and/or justification, but without exploration of alternate explanations or perspectives. Dialogic reflection, in contrast, entails looking for relationships between instances of experience, considering alternate explanations and perspectives, as well as questioning, hypothesizing and interpreting about one's experiences. As such, it bears much semblance to the notions of transformation as enactment and variety. Transformative reflection, in turn, refers to moments when the reflector's original point of view is somehow altered or transformed to take into account the new perspectives they just explored. Finally, critical reflection denotes relating one's experiences to wider socio-cultural and ethical implications. Hence, both transformative and critical reflection constitute examples of personal transformation.

Transformation is often equated with behavioral and/or attitudinal change, as reflected in evaluation metrics. However, these measures do not clearly specify how this change came about or whether it will endure in the long-term. Relative to other UX dimensions, transformation has received relatively little attention to date.

## Evaluation using self-reporting and validated scales

Roth has captured learning and transformation implicitly as an aspect of his granular framework to evaluate the user experience of IDNs (Roth, 2016). The framework has since been explicitly mapped to Murray's category of transformation (Roth & Koenitz, 2016). Specifically, Roth's measurement toolkit captures the dimension of transformation via eudaimonic appreciation.

Which uses a 5-item scale adapting and extending the work of Rowold (2008) and Cupchik and Laszlo (1994). After experiencing an artifact, interactors rate statements on a 5-point Likert scale, ranging from "strongly disagree" to "strongly agree". Typical items are "The experience made me think about my personal situation.", "The experience told me something about life." and "The experience moved me like a piece of art.". The resulting values are aggregated and form a 'transformation score' with 1 being the lowest and 5 being the highest. While these values differ between individuals - based on their experience, expectations, and preferences - the resulting mean score gives an indication of the overall rating of an artifact in regard to its transformation quality. Additionally, a series of qualitative questions asks interactors to elaborate on their experience and the meaning it had to them, allowing for a content analysis to reveal underlying patterns of a meaningful experience.

Understanding the different levels of meaning-making is crucial for the design of transformative experiences. While players interact with an IDN system, they are continuously extracting information to understand past and present events and to plan their actions. The interpretation occurs in terms of narrative game mechanics and their meaning in relation to the narrative theme. Roth, Van Nuenen, and Koenitz (2018) proposed a model to conceptualize the narrative meaning-making processes in the form of a double-hermeneutic circle (cf. Giddens 1987; Veli-Matti Karhulahti, 2012): a) when interacting with the system and b) when interpreting the instantiated narrative at any point of the experience. This model allows the analysis of IDN works as a shared meaning-making activity between designer and audience.

Other works (Mekler et al., 2018) have applied Fleck & Fitzpatrick's reflection framework (2010) to evaluate transformative reflection from qualitative interview data. Via this approach it is possible to (1) identify evidence of reflective thought in interactors' accounts, (2) distinguish it more clearly from non-reflective forms of thinking, as well as (3) differentiate between levels of reflection. However, the effectiveness of this approach is determined by how well interview participants are able to articulate their experience.

## Alternatives to Self-Reporting

To assess the impact of a narrative on users, researchers rely predominantly on self-reported measures, i.e., those in which the user reports directly (through surveys, questionnaires, interviews, etc.) on his or their own experience. The usefulness of these measures is unquestionable; however, they also have important limitations that must be taken into consideration. Self-reported measures assume that the user is able and willing to report on the concept of interest to the researcher. However, depending on various factors (e.g., the topic of the narrative, the context of the measure, the relationship between the user and the researcher) this is not always the case. First, self-reported measures rely on the user's capacity for introspection, which, depending on the psychological variable to be measured, cannot always be taken for granted (Nisbett & Wilson, 1977; Hassin et al., 2004). Moreover, in some cases (e.g., content on sensitive topics) users may be motivated (consciously or unconsciously) to report inaccurate information. Biases such as social desirability bias (Fisher & Katz, 2000), acquiescence bias (Kuru & Pasek, 2016), the bandwagon effect (Schmitt Beck, 2015), or simply careless responding, may affect the validity of reported information.

Different methodological approaches can help to obtain a more complete picture of the effects of exposure to an interactive narrative on users, helping to overcome some of these biases.

One such approach is based on the measurement of physiological variables during exposure to the content. Thus, psychophysiological measures (Potter & Bolls, 2012), such as the measurement of electrodermal activity, heart rate variability, or facial electromyography (the measurement of facial muscle activity), recorded during exposure to content, have proven useful in providing information about attentional and emotional processes. In turn, measures of brain activity, such as functional near-infrared spectroscopy (fNIRS), can help predict behavioral changes in a sample of users after exposure to a message (Burns et al., 2018). Another typology of implicit measures involves the performance of specific tasks after exposure to the message: Implicit Association Tests (Greenwald et al., 2009) or the Implicit Relational Assessment Procedure (Barnes-Holmes et al., 2006) have been used by psychologists to measure attitudes and beliefs of which users are unaware, or are motivated to provide biased self-reports. Finally, another option is the direct measurement of user behavior after interaction with the narrative. This can be done in laboratory studies, for example by offering users the possibility to perform a task that allows them to observe how they behave. This task can range from the possibility of making a donation related to the topic of the narrative to other more creative tasks: for example, Fonseca and Kraus (2016) showed, to different groups of participants, narratives either on the environmental consequences of meat consumption or on non-related topics and, afterwards, they offered snacks to the participants and observed whether they went for vegetarian options.

The use of these indirect techniques presents, of course, significant challenges, including their higher cost in terms of time and money compared to the administration of questionnaires, as well as the need for technical expertise and (e.g., in the case of psychophysiological measures) specific equipment. Therefore, their use should be carefully evaluated in terms of the advantages they can provide in relation to their cost. The use of self-report measures may be advantageous on many occasions (because of their low cost, ease of implementation, and relative ease of interpretation) but, on other occasions, may simply not be effective. Depending on the variable to be considered in the assessment, it is in these cases that alternatives such as the psychophysiological or behavioral measures discussed here may play a key role.

# 04. Outlook: The Future of IDN and Evaluation

Our workgroup looks forward to having user experience measurement from within the experience, while the interactor is still immersed. This will not only take the form of embedded questionnaires but also intra-diegetic measurements that will minimise breakage of immersion and keep the interactor engaged with the narrative.

Another goal is to conceptualise and develop tests to address declarative and procedural knowledge after engaging with an IDN. These tests can be administered directly after the learning experience or with a delay to evaluate long term effects.

Experimenting with and acting within an IDN system allows for declarative and procedural learning. In the procedural system, we acquire knowledge on procedures and skills build usually slowly over time, as we repeatedly get feedback about our actions. This learning is fundamentally different compared to declarative learning, which is about the memorisation of facts and which is faster and requires attention and focus. Evaluation of declarative knowledge is straightforward and can utilize similar tests used in the classroom context. Accessing procedural knowledge is very challenging as it is subconscious and requires procedural performance tests for its evaluation.